\documentclass[12pt, draftclsnofoot, onecolumn]{IEEEtran}
\usepackage{amsmath,amsfonts,amsthm}
\usepackage{algorithmic,algorithm}
\usepackage{array}
\usepackage[caption=false,font=normalsize,labelfont=sf,textfont=sf]{subfig}
\usepackage{textcomp}
\usepackage{stfloats}
\usepackage{url}
\usepackage{verbatim}
\usepackage{graphicx}
\usepackage{cite}
\usepackage{color}
\usepackage[center]{caption}
\begin{document}

\title{Learning-Based Joint Beamforming and Antenna Movement Design for Movable Antenna Systems}

%\title{Learn to Beamform and Track Double Side Antenna Movement for Movable Antenna System}

\author{Caihao Weng,~Yuanbin Chen,~Lipeng~Zhu,~\IEEEmembership{Member,~IEEE,}~and~Ying Wang,~\IEEEmembership{Member,~IEEE}
        % <-this % stops a space
\thanks{C. Weng, Y. Chen and Y. Wang are with the State Key Laboratory of Networking and Switching Technology, Beijing University of Posts and Telecommunications, Beijing, China 100876 (e-mail: wengcaihao@bupt.edu.cn; chen\_yuanbin@163.com; wangying@bupt.edu.cn).

L. Zhu is with the Department of Electrical and Computer
Engineering, National University of Singapore, Singapore 117583 (e-mail: zhulp@nus.edu.sg).

}
\vspace{-2em}
}


\maketitle

\begin{abstract}
In this paper, we investigate a multi-receiver communication system enabled by movable antennas (MAs). Specifically, the transmit beamforming and the double-side (DS) antenna movement at the transceiver are jointly designed to maximize the sum-rate of all receivers under imperfect channel state information (CSI). 
% {\color{red}Since imperfect CSI is a more practical channel model compared with perfect CSI due to the inevitable channel estimation errors (CEEs). }
% Nevertheless, the MA movement is subject to complex constraints and its optimization is coupled with the CSI, rendering the formulated problem intractable to solve.
Since the formulated problem is non-convex with highly coupled variables, conventional optimization methods cannot solve it efficiently.
To address these challenges, an effective learning-based algorithm is proposed, namely heterogeneous multi-agent deep deterministic policy gradient (MADDPG), which incorporates two agents to learn policies for beamforming and movement of MAs, respectively. Based on the offline learning under numerous imperfect CSI, the proposed heterogeneous MADDPG can output the solutions for transmit beamforming and antenna movement in real time. Simulation results validate the effectiveness of the proposed algorithm, and the MA can significantly improve the sum-rate performance of multiple receivers compared to other benchmark schemes.
\end{abstract}


\section{Introduction}
To meet the dramatically growing demand of wireless applications, wireless communication has imposed stringent requirements on capacity enhancement. By using multiple antennas to transmit independent data streams to harvest spatial multiplexing gain, multiple-input multiple-output (MIMO) and massive MIMO have become the key enabling technologies to improve the transmission rate in the fifth-generation (5G) mobile communications and beyond. However, they generally incur significant hardware costs and energy consumption due to the integration of an increasing number of antennas and radio frequency (RF) chains for wireless systems operating at higher frequency bands\cite{mimo}.

In order to decrease the number of RF chains, antenna selection (AS) is an effective method that generates favorable channels by selecting a small number of activated antennas from a large number of candidate antennas, which utilizes the spatial degrees of freedom (DoFs) of wireless channels to improve communication performance \cite{AS}. 
%Nevertheless, the presence of numerous candidate antennas escalates the cost and computational complexity of the AS technology. {\color{red}To reduce the number of antennas, utilize the spatial DoFs and improve system performance, fluid antenna system (FAS) was proposed.} 
However, the inclusion of numerous candidate antennas increases both the hardware cost and computational complexity associated with AS technology. To address this issue, fluid antenna system (FAS) was proposed to deploy liquid antennas or pixel-based antennas to exploit the spatial DoFs for improving system performance ~\cite{FAS}. This antenna can be flexibly switched among multiple discrete and fixed ports in a one/two-dimensional (1D/2D) space, enabling the FAS to capture the strongest received signal. Nonetheless, both MIMO/massive MIMO with fixed antenna positions and discrete-port based FAS in existing works cannot fully exploit the spatial variation of wireless channels, especially for a large antenna/port distance.

To exploit more spatial DoFs, movable antenna (MA) has been incorporated into wireless communication systems~\cite{MA-Mag}. By employing flexible cables to connect the RF chains to the MAs, the positions of MAs can be flexibly/continuously adjusted in three-dimensional (3D) space with the aid of drivers, which can significantly reconfigure the wireless channel between the transmitter and the receiver and increase the communication performance in terms of spatial diversity, spatial multiplexing, and flexible beamforming. However, since the antenna movement requires sufficient time, it is difficult for MA systems to achieve complete performance gains under rapidly changing wireless channels. Therefore, MA is typically applied in scenarios where the wireless channels (or dominant channel paths) change slowly over time, such as that with fixed or slow-moving transceiver locations\cite{TWC-2}.

Existing studies have validated the significant performance improvements of MA-enabled wireless systems\cite{TWC-2,TWC-capacity,multiuser-1}. Specifically, the field-response channel model was proposed in \cite{TWC-2} to characterize the wireless channel variation with respect to the transmit and receive MAs' positions. In \cite{TWC-capacity}, the capacity of the MA-enabled MIMO system was maximized, in which an alternating optimization algorithm is proposed for designing the position of each MA and the transmit covariance matrix. Furthermore, MA-aided multi-user communication was studied in \cite{multiuser-1,multiuser-2,multiuser-3,multiuser-4} to minimize the total transmit power of users subject to a minimum rate requirement or maximize the sum rate of users subject to the total power budget.
% In \cite{multiuser-2}, an optimization problem is formulated to maximize the minimum achievable rate of MA-enable multi-user communications by jointly optimizing the location of the MAs, receive combing at the BS, and the transmit power of users, a particle swarm optimization (PSO) method was proposed to solve this challenging non-convex optimization problem. 
However, existing studies on MA-aided multi-user communications mainly consider the single-side antenna movement design\cite{multiuser-1,multiuser-2,multiuser-3,multiuser-4}, which cannot to fully utilize the spatial DoFs of both the transmitter and receiver. Therefore, there are still considerable potentials for system performance improvement. Moreover, existing works generally rely on traditional optimization algorithms for MA position, which not only incur high computational complexities but also limit practical applications with imperfect channel state information (CSI). In contrast, model-free machine learning method can efficiently reduce the online computational complexity while achieving performance comparable to traditional approaches.

Motivated by these considerations, we investigate in this paper MA-enabled multi-receiver communication systems based on unsupervised deep reinforcement leaning (DRL). Specifically, the transmit beamforming and the double-side (DS) antenna movement at the transceivers are jointly designed. First, to enhance the system capacity, we study a sum-rate maximization problem under imperfect CSI. Then, to solve this problem, a heterogeneous multi-agent deep deterministic policy gradient (MADDPG) algorithm is proposed. In particular, we divide the agents into the beamforming agents and the MA agents, enabling them to learn beamforming policy and antenna movement policy, respectively.
% This heterogeneous architecture, as opposed to treating the transmitter and receiver as individual agents, avoids relying on outdated CSI learning beamforming policy, leading to more significant performance improvements. 
Based on the offline learning under numerous imperfect CSI, the proposed MADDPG can output the solutions for transmit beamforming and antenna movement in real time. Finally, simulation results are provided to demonstrate the effectiveness of the proposed heterogeneous MADDPG algorithm in terms of increasing system sum-rate compared to other benchmark schemes.

\section{SYSTEM MODEL}
\begin{figure*}[!t]
\centering
\includegraphics[width=0.8\linewidth]{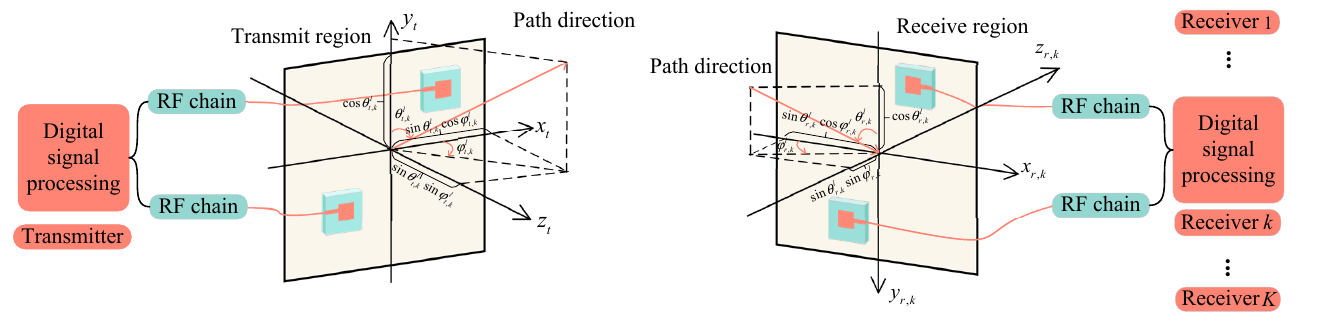}
\caption{MA-enabled multi-receiver communication system.}
\label{fig1}
\vspace{-1em}
\end{figure*}
\subsection{Scenario}
In this paper, we consider a time-division duplexing (TDD)-based MIMO communication system as shown in Fig.~\ref{fig1}. 
% The transmitter is equipped with $N$ MAs and $N_{RF}^t$ RF chains to serve $K$ receivers, where the $k$-th receiver is equipped with $M_k$ MAs and $N_{RF}^k$ RF chains. For simplicity, we assume that there is only one stream between the transmitter and each receiver\cite{single-stream}. Hence, the number of RF chains is constrained by $K \le N_{RF}^t \le N$ for the transmitter and $1\le N_{RF}^k \le M_{k}$ for receiver $k$.
The transmitter is equipped with $N$ MAs to serve $K$ receivers, where the $k$-th receiver is equipped with $M_k$ MAs.
% The transmit and receive MAs are connected to the RF chain through flexible cables, allowing real time adjustment of their positions\cite{TWC-capacity}. 
$\mathcal{R}_T$ and $\mathcal{R}_{R,k}$ denote the predefined two-dimensional (2D) regions where the transmit and receive MAs can move freely, respectively. Without loss of generality, we assume that the $\mathcal{R}_{T}$ and $\mathcal{R}_{R,k}$ are square regions with size $A\times A$ in the $x_t$-$y_t$ and $x_{r,k}$-$y_{r,k}$ planes, respectively.

% and the transmit region and the $k$-th receive region are centered at the origin $\mathbf{c}_o=[0,0,0]^T$ and $\mathbf{c}_{o,k}=[x_{o,k},y_{o,k},z_{o,k}]^T$, respectively. 
The Cartesian coordinate of the $n$-th transmit MA is $\mathbf{c}_n=[x_n,y_n]^T\in\mathcal{R}_T$. Denote the collection of the coordinates of $N$ transmit MAs by $\mathcal{C}_T=[\mathbf{c}_1,\ldots,\mathbf{c}_N ]\in \mathbb{R}^{2\times N}$. Similarly, the Cartesian coordinate of the $m$-th MA of the $k$-th receiver is $\mathbf{c}_{k,m}=[x_{k,m},y_{k,m}]^T \in \mathcal{R}_{R,k}$. Denote the collection of the coordinates of the $K$ receivers' MAs by $\mathcal{C}_R=[\mathcal{C}_{R,1},\mathcal{C}_{R,2},\ldots,\mathcal{C}_{R,K}]$, where $\mathcal{C}_{R,k}=[\mathbf{c}_{k,1},\ldots,\mathbf{c}_{k,M_k}]\in\mathbb{R}^{2\times M_k}$ is the collection of the coordinates of the $M_k$ receive MAs at receiver $k$.

This paper considers the far-field wireless channel with $L_k$ paths between the transmitter and the $k$-th receiver. Therefore, the angle of arrival (AoA) and the angle of departure (AoD) for each channel path component remain unchanged for MAs at different locations. Denote the elevation and azimuth AoDs of the $l$-th $(l=1,2,\ldots,L_k)$ path between the transmitter and the $k$-th receiver as $\theta^l_{t,k}\in[0,\pi]$ and $\phi^l_{t,k}\in[0,\pi]$. Similarly, $\theta^l_{r,k}\in [0,\pi]$ and $\phi^l_{r,k}\in [0,\pi]$ denote the elevation and azimuth AoAs of the $l$-th path between the transmitter and the $k$-th receiver. Specifically, $l=1$ denotes the line-of-sight (LoS) path, while $l>1$ indicates the non-line-of-sight (NLoS) path.

\subsection{Channel Model and Communication Protocol}
For the $l$-th path between the transmitter and the $k$-th receiver, the normalized wave vector can be expressed as $\mathbf{k}^l_{t,k}=[\sin\theta^l_{t,k}\cos\phi^l_{t,k},\cos\theta^l_{t,k},\sin\theta^l_{t,k}\sin\phi^l_{t,k} ]^T$. 
% Let $\mathbf{c}_t=[x_t,y_t,0]^T$ and $\mathbf{c}_o=[0,0,0]^T$ denote the three-dimensional (3D) coordinate vectors of the transmit MA and the origin of the transmit region.
Therefore, the propagation difference between the $n$-th transmit MA and the origin of the transmit region is
\begin{equation}
d^l_{t,k,n} = x_{n}\sin\theta^l_{t,k}\cos\phi^l_{t,k}+y_{n}\cos\theta^l_{t,k}.
\end{equation}
Then, the propagation phase difference between them can be expressed as $2\pi d^l_{t,k}/\lambda$, where $\lambda$ is the wavelength of the operating frequency. Thus, the transmit array response vector of the $l$-th path of the $k$-th receiver $\mathbf{a}(\mathbf{d}^l_{t,k})\in\mathbb{C}^{1\times N}$ is expressed as
\begin{equation}
\mathbf{a}\left(\mathbf{d}^l_{t,k}\right)=\left[e^{j\frac{2\pi}{\lambda}d^l_{t,k,1}},e^{j\frac{2\pi}{\lambda}d^l_{t,k,2}},\ldots,e^{j\frac{2\pi}{\lambda}d^l_{t,k,N}}\right],
\end{equation}
with $\mathbf{d}^l_{t,k}=[d^l_{t,k,1},d^l_{t,k,2},\ldots,d^l_{t,k,N}]\in\mathbb{C}^{1\times N}$.
% and $d^l_{t,k,n}$ is the propagation difference between the $n$-th transmit MA and the origin of the transmit region.
Similarly, the receive array response vector of the $l$-th path of the $k$-th receiver $\mathbf{f}(\mathbf{d}^l_{r,k})\in\mathbb{C}^{1\times M_k}$ is expressed as
\begin{equation}
\mathbf{f}\left(\mathbf{d}^l_{r,k}\right)=\left[e^{j\frac{2\pi}{\lambda}d^l_{r,k,1}},e^{j\frac{2\pi}{\lambda}d^l_{r,k,2}},\ldots,e^{j\frac{2\pi}{\lambda}d^l_{r,k,M_k}}\right],
\end{equation}
where $\mathbf{d}^l_{r,k}=[d^l_{r,k,1},d^l_{r,k,2},\ldots,d^l_{r,k,M_k}]\in\mathbb{C}^{1\times M_k}$ and $d^l_{r,k,m}=x_{k,m}sin\theta^l_{r,k}cos\phi^l_{r,k}+y_{k,m}\cos\theta^l_{r,k}$ is the propagation difference between the $m$-th MA of the $k$-th receiver and the origin of the receive region for the $l$-th path.

Therefore, the perfect/actual channel matrix between the transmitter and the $k$-th receiver is expressed as
\begin{equation}
\mathbf{H}_k=\sum_{l=1}^Lg_{l,k}\mathbf{a}^{H}(\mathbf{d}_{t,k}^l)\mathbf{f}(\mathbf{d}_{r,k}^l),
\end{equation}
% where $g_{1,k}^{{\prime}}=\sqrt{\frac{\varepsilon}{1+\varepsilon}}g_{1,k}$, $g_{l,k}^{{\prime}}=\sqrt{\frac{1}{1+\varepsilon}}g_{l,k},l>1$, $\varepsilon$ and $g_{l,k}$ are the Rician factor and the channel gain of the $l$-th path of the $k$-th receiver, respectively.
where $g_{l,k}$ is the channel gain of the $l$-th path of the $k$-th receiver.

For simplicity, we assume that there is only one data stream between the transmitter and each receiver\cite{single-stream}. Let $x_k\in\mathbb{C}$ denote the transmit symbol intended for receiver $k$ with $\mathbb{E}[x_k x^H_k]=1$ and $\mathbf{w}_k\in\mathbb{C}^{N\times 1}$ denote its transmit beamforming vector. Thus, the received signal of the $k$-th receiver under perfect CSI is expressed as

\begin{equation}
\mathbf{y}_k=\mathbf{H}_k^H\mathbf{w}_kx_k+\sum_{k^{\prime}\neq k}\mathbf{H}_k^H\mathbf{w}_{k^{\prime}}x_{k^{\prime}}+\mathbf{n}_k,
\end{equation}
where $\mathbf{n}_k\in\mathbb{C}^{M_k\times1}$ is the additive white Gaussian noise
(AWGN) with power $\sigma^2$, i.e., $\mathbf{n}_k\sim\mathcal{CN}\left(\mathbf{0}_{M_k},\sigma^2 \mathbf{I}_{M_k} \right).$
Then, the achievable rate of the $k$-th receiver under perfect CSI can be expressed as
\begin{equation}
R_k^{\textit{p}}=\log_2\left(1+\mathbf{w}_k^H\mathbf{H}_k\mathbf{J}_{p,k}^{-1}\mathbf{H}_k^H\mathbf{w}_k\right),
\end{equation}
where $\mathbf{J}_{p,k}=\sum_{k^{\prime}\neq k}\mathbf{H}_k^H\mathbf{w}_{k^{\prime}}\mathbf{w}_{k^{\prime}}^H\mathbf{H}_k+\sigma^2\mathbf{I}_{M_k}$ is the covariance matrix of interference plus noise at receiver $k$.

By utilizing the existing channel estimation method for MA systems \cite{CE}, we can obtain the CSI between the transmitter and receivers. However, there inevitably exist channel estimation errors (CEEs) between the perfect/actual channel matrix $\mathbf{H}_k$ and the imperfect/estimated channel matrix $\hat{\mathbf{H}}_k$, which may erode the system performance. For notation simplicity, we define $\hat{\mathbf{H}}=\{\hat{\mathbf{H}}_1,\hat{\mathbf{H}}_2,\ldots,\hat{\mathbf{H}}_K\}$. Therefore, considering CEEs $\Delta\mathbf{H}_k$, the actual CSI can be expressed as 

\begin{equation}
\mathbf{H}_k=\hat{\mathbf{H}}_k+\Delta\mathbf{H}_k.
\end{equation}

According to \cite{CEEs-2}, CEEs can be model as circularly symmetric complex Gaussian (CSCG) variables, i.e., $\Delta\mathbf{H}_k\sim\mathcal{CN}(\mathbf{0},\mathbf{A}_k\otimes\mathbf{B}_k)$, where $\mathbf{A}_k$ and $\mathbf{B}_k$ are the channel estimation error covariance matrices at the receiver side and the transmitter side, respectively\cite{CEEs}.
% Generally speaking, CEEs are related to the employed channel estimation method.
Assuming that CEEs are independent of the estimated channel matrices and their statistics are known to the transceivers, we thus have
\begin{equation}
\mathbf{H}_k\sim\mathcal{CN}(\hat{\mathbf{H}}_k,\mathbf{A}_k\otimes\mathbf{B}_k).
\end{equation}

Due to the existence of CEEs, the received signal at receiver $k$ under imperfect CSI is expressed as
% \begin{equation}
% \mathbf{y}_{k}=\mathbf{H}_{k}^{H}\mathbf{w}_{k}x_{k}+\underbrace{\sum_{k}\Delta\mathbf{H}_{k}^{H}\mathbf{w}_{k}x_{k}}_{\mathrm{CEEs}}+\underbrace{\sum_{k^{\prime}\neq k}\mathbf{H}_{k^{\prime}}^{H}\mathbf{w}_{k^{\prime}}x_{k^{\prime}}+\mathbf{n}_{k}}_{\text{interference signal and noise}}.
% \end{equation}
\begin{equation}
\mathbf{y}_{k}=\hat{\mathbf{H}}_{k}^{H}\mathbf{w}_{k}x_{k}+\sum_{k^{\prime}}\Delta\mathbf{H}_{k}^{H}\mathbf{w}_{k^{\prime}}x_{k^{\prime}}+\sum_{k^{\prime}\neq k}\hat{\mathbf{H}}_{k}^{H}\mathbf{w}_{k^{\prime}}x_{k^{\prime}}+\mathbf{n}_{k}.
\end{equation}
Then, the expectation of the achievable rate of the $k$-th receiver under imperfect CSI can be expressed as
\begin{equation}
R_k^{i} = \mathbb{E}_{\left\{\Delta\mathbf{H}_k\right\}}\left\{\log_2\left(1+\mathbf{w}_k^H\hat{\mathbf{H}}_k\hat{\mathbf{J}}_{i,k}^{-1}\hat{\mathbf{H}}_k^H\mathbf{w}_k\right)\right\},
\end{equation}
where $\hat{\mathbf{J}}_{i,k} = \sum_{k^{\prime}}{\Delta\mathbf{H}_{k}^H\mathbf{w}_{k^{\prime}}\mathbf{w}_{k^{\prime}}^H\Delta\mathbf{H}_k}$$ + \sum_{k^{\prime}\neq k}{ \hat{\mathbf{H}}_k^H\mathbf{w}_{k^{\prime}}\mathbf{w}_{k^{\prime}}^H\hat{\mathbf{H}}_k}$
$+ \sigma^2\mathbf{I}$ is the covariance matrix of interference plus noise plus CEEs at receiver $k$. 
\subsection{Problem Formulation}
We aim to improve the sum-rate under imperfect CSI by jointly optimizing the transmit beamforming vectors $\mathbf{W}=[\mathbf{w}_1,\mathbf{w}_2,...,\mathbf{w}_K]\in\mathbb{C}^{N\times K}$ and the DS antenna positions $\mathcal{C}_T,\mathcal{C}_R$. The optimization problem is formulated as
\begin{subequations}
\begin{align}
(\mathrm{P1})\quad\max_{\mathcal{C}_{T},\mathcal{C}_{R},\mathbf{W}} & \sum_{k=1}^{K}R_{k}^i \label{Za} \\
\mathrm{s.t.}\quad&\sum_{k=1}^{K}tr(\mathbf{w}_{k}\mathbf{w}_{k}^{H})\leq P \\
&\|\mathbf{c}_{n}-\mathbf{c}_{n^{\prime}}\|\geq\lambda/2,n\neq n^{\prime} \\
&\|\mathbf{c}_{k,m}-\mathbf{c}_{k,m^{\prime}}\|\geq\lambda/2,m\neq m^{\prime},\forall k \\
&\mathbf{c}_{n}\in\mathcal{R}_{T},\forall n \\
&\mathbf{c}_{k,m}\in\mathcal{R}_{R,k},\forall m,k,
\end{align}
\end{subequations}
% where $R_k=\{R_k^{\textit{p}} , R_k^{i} \}$ and 
where $P$ is the power budget. The minimum distance constraints between MAs in (11c) and (11d) avoids the coupling effect between MAs.

Problem (P1) is difficult to solve because $R^{i}_k$ has no explicit expression. In the following theorem, we will derive the closed-form upper bound on the rate of receiver $k$, then we will maximize the upper bound by jointly optimizing the transmit beamforming vector and the DS antenna movement.

\noindent\textbf{Theorem 1}: In MA-enabled MIMO systems, the achievable rate of receiver $k$ can be upper bounded by
\begin{equation}
{R}_k^{\textit{i}}\le\hat{R}_k^{\textit{i}} = \log_2\left(1+\mathbf{w}_k^H\hat{\mathbf{H}}_k\mathbf{U}_k^{-1}\hat{\mathbf{H}}_k^H\mathbf{w}_k\right),
\end{equation}
where $\mathbf{U}_k=\sum_ktr\left(\mathbf{B}_k\mathbf{w}_k\mathbf{w}_k^H\right)\mathbf{A}_k^T$$+\sum_{k^{\prime}\neq k}\hat{\mathbf{H}}_k^H\mathbf{w}_{k^{\prime}}\mathbf{w}_{k^{\prime}}^H\hat{\mathbf{H}}_k$ $+\sigma_k^2\mathbf{I}$.

\begin{proof}
Based on Jensen’s inequality, we have
\begin{equation}
\begin{aligned}
\mathbb{E}_{\left\{\Delta\mathbf{H}_k\right\}} \left\{\log_2\left(1+\mathbf{w}_k^H\hat{\mathbf{H}}_k\hat{\mathbf{J}}_{i,k}^{-1}\hat{\mathbf{H}}_k^H\mathbf{w}_k\right)\right\} \leq\log_2\det\left(1+\mathbf{w}_k^H\hat{\mathbf{H}}_k\mathbb{E}_{\left\{\Delta\mathbf{H}_k\right\}}\left\{\hat{\mathbf{J}}_{i,k}\right\}^{-1}\hat{\mathbf{H}}_k^H\mathbf{w}_k\right),
\end{aligned}
\end{equation}
where $\mathbb{E}_{\left\{\Delta\mathbf{H}_k\right\}}\left\{\hat{\mathbf{J}}_{i,k}\right\}$ can be calculated as

\begin{equation}
\begin{aligned}
\mathbf{U}_k=\mathbb{E}_{\left\{\Delta\mathbf{H}_k\right\}}\left\{\hat{\mathbf{J}}_{i,k}\right\} 
\overset{(a)}{\operatorname*{=}}\sum_{k^{\prime}}tr\left(\mathbf{B}_k\mathbf{w}_k\mathbf{w}_k^H\right)\mathbf{A}_k^T +\sum_{k^{\prime}\neq k}\hat{\mathbf{H}}_k^H\mathbf{w}_{k^{\prime}}\mathbf{w}_{k^{\prime}}^H\hat{\mathbf{H}}_k+\sigma_k^2\mathbf{I},
\end{aligned}
\end{equation}
where (a) holds because for $\Delta\mathbf{H}\sim\mathcal{CN}\left(0,\mathbf{A}\otimes\mathbf{B}\right)$, there is $\mathbb{E}_{\{\Delta\mathbf{H}\}}\left[{\Delta\mathbf{H}}^H\mathbf{w}_{_k}\mathbf{w}_{_k}^H\Delta\mathbf{H}\right]=tr(\mathbf{B}\mathbf{w}_{_k}$
$\mathbf{w}_{_k}^H)\mathbf{A}^T$~\cite{math}.
\end{proof}

According to \noindent\textbf{Theorem 1}, problem (P1) can be recast to
\begin{equation}
\begin{aligned}
(\mathrm{P2})\quad\max_{\mathcal{C}_{T},\mathcal{C}_{R},\mathbf{W}} & \sum_{k=1}^{K}\hat{R}^i_k 
 \\
\mathrm{s.t.}\quad& \mathrm{(11b) - (11f)}.
\end{aligned}
\end{equation}
% where $\hat{R}_k=\{R_k^{p},\hat{R}_k^{i}\}$.

Problem (P2) is still difficult to solve because it is a non-convex problem. Specifically, the objective function is a non-convex function over antenna positions and the constraints (11c) and (11d) are also non-convex. In the following section, a DRL based method is proposed to solve problem (P2) by jointly optimizing the transmit beamforming vector and the DS antenna movement.

We propose an offline-trained MADDPG algorithm to resolve problem (P2), which empower agents to autonomously learn optimal policies for both beamforming and the movement of MAs \cite{MADDPG}. For the considered MA system, a straightforward idea involves designating each transmitter and receiver as an independent agent, also termed as the TR-MADDPG. 
% However, this approach encounters significant challenges. Assigning the transmitter the dual responsibilities of determining beamforming strategies and MA movement complicates the learning process. Specifically, the movement of an MA changes the channel matrix, consequently rendering the beamforming policy predicated on previously estimated CSI. This obsolescence not only degrades system performance but also expands the action space dimension for the transmitter agent to an impractical dimension, complicating the policy learning process.
However, if the transmitter is regarded as an agent responsible for learning policies for beamforming and movement of MAs simultaneously, the movement of MA will reconfigure the channel matrix, which leads to the learn of beamforming policy using outdated CSI, degrading system performance. In addition, the action space dimension of the transmitter agent will become excessively large and it is difficult to learn the policies.

\begin{figure}[!t]
\centering
\includegraphics[width=0.8\linewidth]{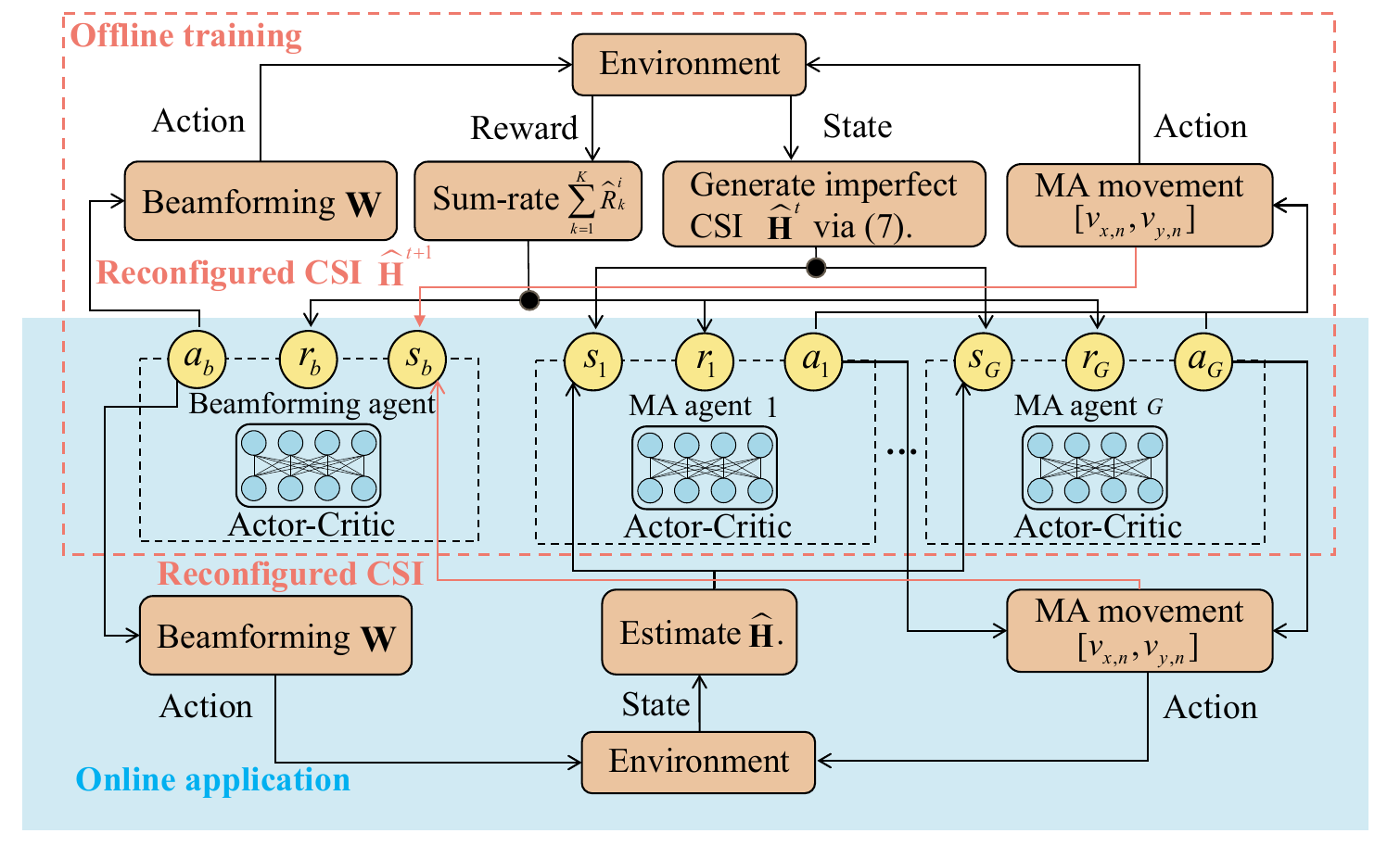}
\caption{The framework of proposed heterogeneous MADDPG.}
\label{fig2}
\vspace{-1em}
\end{figure}

To tackle these challenges, a heterogeneous MADDPG framework for the MA system is proposed, as illustrated in Fig.~\ref{fig2}. The framework involves two types of agents, where the beamforming agent is responsible for learning the beamforming policy of the transmitter and the MA agent is responsible for learning the antenna movement policy of the transceivers. The detailed process of heterogeneous MADDPG is described in \textbf{Algorithm 1}. The proposed MADDPG comprises a single beamforming agent and $G$ MA agents, with $G=K+1$. To maximize the sum-rate under imperfect CSI, the states, actions and rewards of the two agents are defined as follows.

\section{Heterogeneous MADDPG-Based Solution}

\subsection{MA Agent}
\subsubsection{State $s_{t,g}$} 

In the $t$-th training time slot, the state of the MA agent contains the imperfect CSI between the transmitter and receivers. Specifically, for the MA agent corresponding to the transmitter, the state incorporates the CSI from the transmitter to all receivers, i.e., $\hat{\mathbf{H}}^t$. While for the MA agent corresponding to receiver $k$, the state can be defined by the CSI between the transmitter and the $k$-th receiver, i.e., $\hat{\mathbf{H}}^t_k$. Furthermore, the proposed algorithm consists of two stages, i.e., the offline training and the online application. In the offline training stage, we first generate perfect CSI based on (4), and then obtain imperfect CSI according to (7) and CEEs $\Delta\mathbf{H}_k$. The update of model parameters are based on the imperfect CSI. In the online application, the proposed algorithm can directly output favorable actions based on the estimated imperfect CSI.

\subsubsection{Action $a_{t,g}$} We define the MAs' velocity projections on the $x$-axis and the $y$-axis as action. For example, the action of the MA agent corresponding to the transmitter is $[v_{x,1},v_{y,1},\ldots,v_{x,n},$
$v_{y,n},\ldots,v_{x,N},y_{y,N}]$, where $v_{x,n}$ and $v_{y,n}$ are the velocities along the $x$-axis and the $y$-axis of the $n$-th MA, respectively..

\subsubsection{Reward $r_{t,g}$} The reward of the $g$-th MA agent is
\begin{equation}
\begin{aligned}
r_{t,g}=\sum_{k=1}^{K}\hat{R}^i_{k}-c_{1}R_{blp}^{g}-c_{2}R_{msp}^{g},
\end{aligned}
\end{equation}
where $R_{blp}^g$ and $R_{msp}^g$ are the penalties when the constraints (11c)-(11d) and (11e)-(11f) are not satisfied. The coefficients $c_1$ and $c_2$ are the weights used to balance the penalties and the sum-rate.

\subsection{Beamforming Agent}  
\subsubsection{State $s_{t,b}$} In the $t$-th traing time slot, the state of the MA agent is the CSI between the transmitter and the receiver. 
As shown in Fig.\ref{fig2}, the state of the beamforming agent $s_{t,b}$ is the reconfigured CSI $\hat{\mathbf{H}}^{t+1}$, because the movement of MA reconfigures the CSI, making $\hat{\mathbf{H}}^t$ outdated.

\subsubsection{Action $a_{t,b}$} We define the real part and the imaginary part of the transmit beamforming vector as action, i.e., $\operatorname{Re}\{\mathbf{W}\}$ and $\operatorname{Im}\{\mathbf{W}\}$.

\subsubsection{Reward $r_{t,b}$} The reward of the beamforming agent is
\begin{equation}
r_{t,b}=\sum_{k=1}^K\hat{R}^i_{k}-c_3R_{pp},
\end{equation}
where $R_{pp}$ is the penalty when the constraint (11b) is not satisfied and the coefficient $c_3$ is the weight used to balance the penalty and the sum-rate.

\begin{algorithm}[t]
\small
	\caption{Offline Heterogeneous MADDPG Algorithm}
	\label{alg1}
	\begin{algorithmic}[1]
		\STATE Initialize the actor and critic networks, target actor and critic networks;
            \FOR{Episode $n_{epi}=1,2,\ldots,N_{epi}$}
                \STATE Initialize the positions of MAs;
                \FOR{Time slot $t=1,2,\ldots,N_{step}$}
                    \STATE Calculate perfect CSI $\mathbf{H}_k$ according to (4);
                    \STATE Get imperfect CSI $\hat{\mathbf{H}},\hat{\mathbf{H}}_1,\hat{\mathbf{H}}_2,\ldots,\hat{\mathbf{H}}_K$ as observations $s_{t,1},s_{t,2},\ldots,s_{t,G}$ according to (7);
                    \STATE MA agents select actions $a_{t,1},a_{t,2},\ldots,a_{t,G}$;
                    \STATE MA agents execute actions $a_{t,1},a_{t,2},\ldots,a_{t,G}$, receive rewards $r_{t,1},r_{t,2},\ldots,r_{t,G}$ and new states $s_{t+1,1},s_{t+1,2},\ldots,s_{t+1,G}$;
                    \STATE Beamforming agent observes $s_{t+1,1}$ as $s_{t,b}$, i.e, reconfigured CSI $\hat{\mathbf{H}}^{t+1}$, and then select action $a_{t,b}$;
                    \STATE Beamforming agent executes action $a_{t,b}$, and receive reward $r_{t,b}$;
                    \STATE Store the transitions $\left[s_{t,g},a_{t,g},r_{t,g},s_{t+1,g} \right]$ and $\left[s_{t,b},a_{t,b},r_{t,b},s_{t+1,b} \right]$ into the memory queues;
                    \STATE Sample a random mini bath to update actor and critic networks;
                    \STATE Update target actor and critic networks.
                \ENDFOR
            \ENDFOR
	\end{algorithmic}  
\end{algorithm}

% {\color{red}
% \subsection{Complexity Analysis}
% This subsection discusses the time complexity of the proposed algorithm. the main factor influencing the computational complexity is the dimension of the actor network and the critic network. Specifically, the computational complexity required for training a actor network using a period of experience is $\mathcal{O}_{actor}=\mathcal{O}(CE+E^{2}+ET)$, where $C$ and $E$ are the dimension of the state space and the action space, $E$ is number of neurons of the hidden layer. Similarly, the computational complexity of training a critic network is $\mathcal{O}_{critic}=\mathcal{O}((C+T)E+E^{2}+E)$. Then, the computational of training an agent is ${\mathcal O}_{agent}={\mathcal O}(2(CE+E^{2}+ET+(C+T)E+E^{2}+E))$. Thus, the computational complexity of the proposed heterogeneous MADDPG is $\mathcal{O}=\mathcal{O}(2(G+1)B(CE+E^2+ET+(C+T)E+E^2+E))$, where $B$ is the number of experiences in a sample batch.}

\section{Simulation Results}

\begin{figure*}[htbp]
\begin{minipage}[t]{0.33\textwidth}
\centering
\includegraphics[scale=0.33]{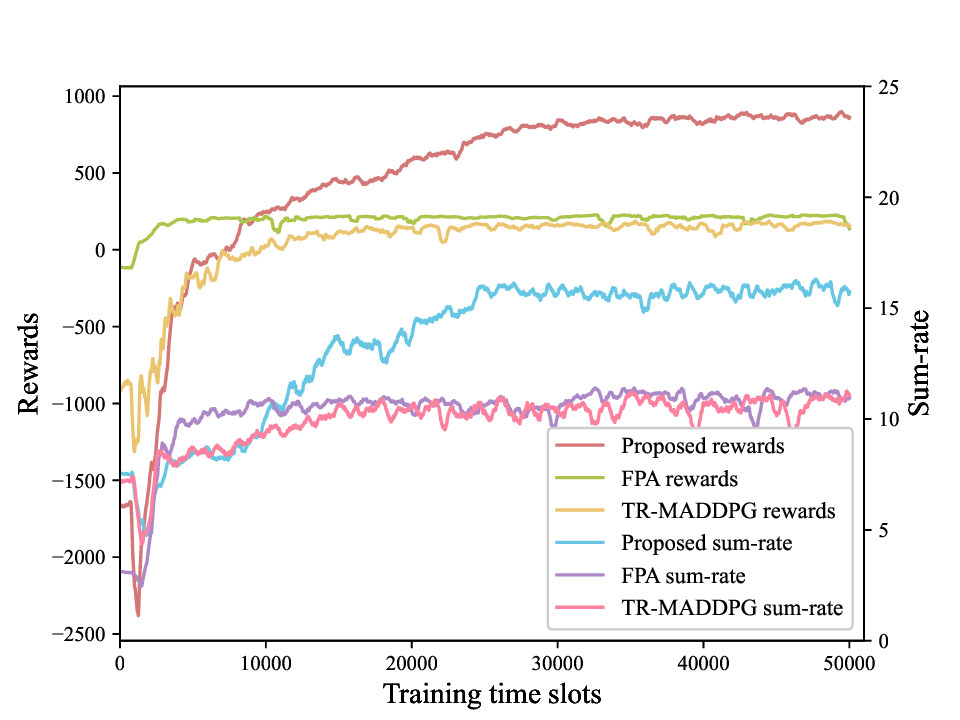}
\caption{Reward and sum-rate performance versus training time slots.}
\label{fig3}
\end{minipage}
\begin{minipage}[t]{0.33\textwidth}
\centering
\includegraphics[scale=0.33]{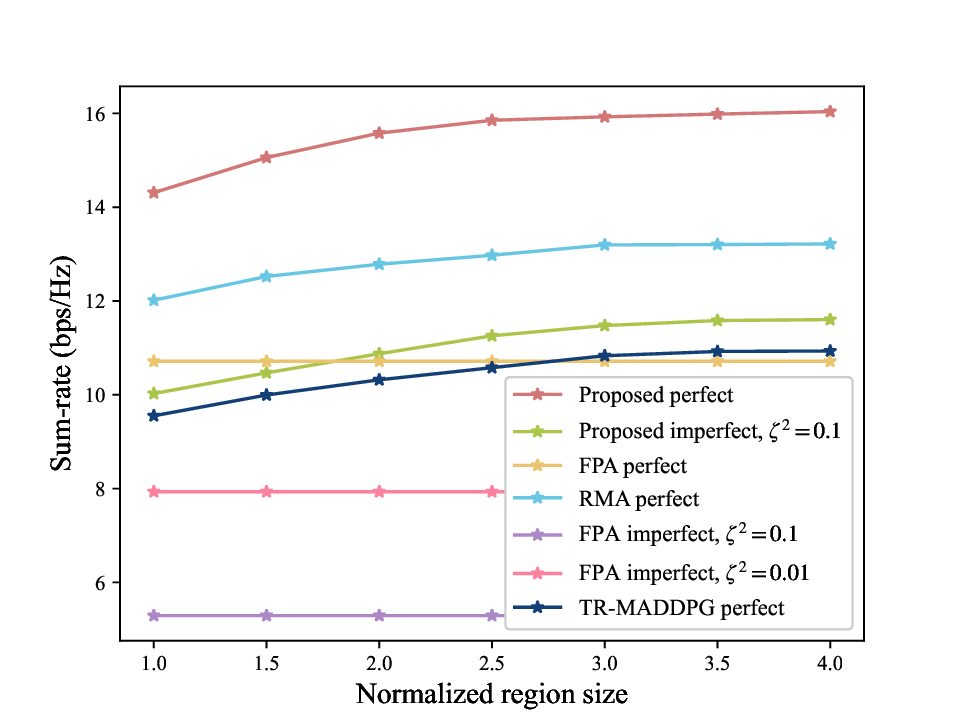}
\caption{Sum-rate performance versus normalized region size.}
\label{fig4}
\end{minipage}
\begin{minipage}[t]{0.32\textwidth}
\centering
\includegraphics[scale=0.33]{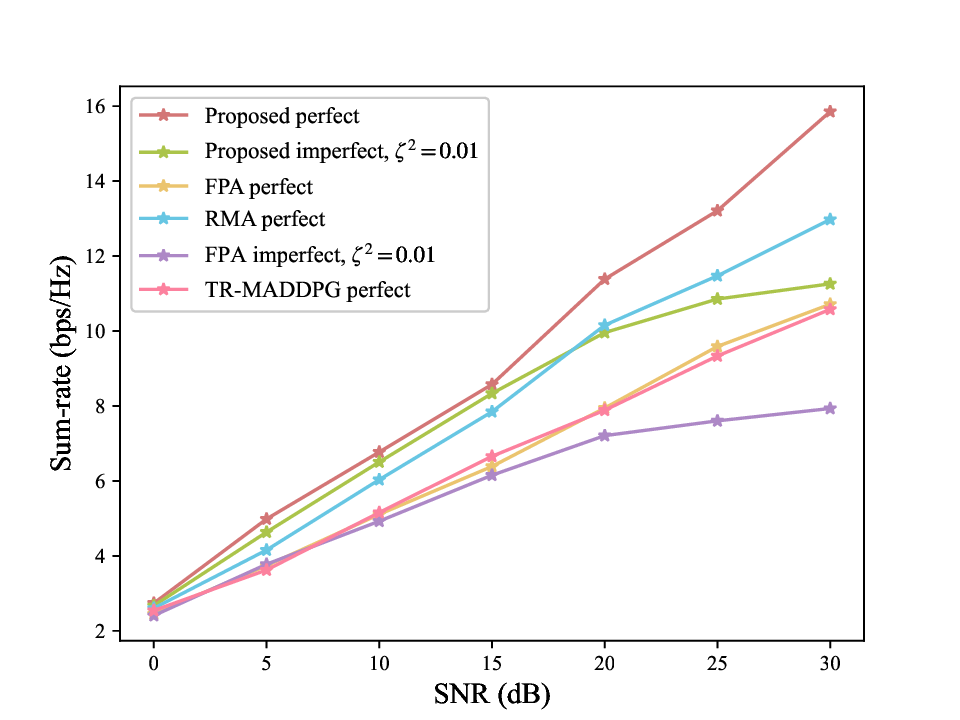}
\caption{Sum-rate performance versus SNR.}
\label{fig5}
\end{minipage}
\vspace{-1em}
\end{figure*}

This section presents the simulation results to evaluate the performance of the proposed heterogeneous MADDPG. 
In the simulation, we consider a MIMO system with one transmitter serving two receivers.
Two MAs are configured at both the transmitter and receivers.
% and the size of the transmit and receiver region is both $A\times A$. 
We consider a geometric channel model with 3 channel paths between the transmitter and each receiver, including one LoS path and two NLoS paths. The transmitter and receivers are located at fixed positions, where the AoAs and AoDs for the NLoS paths are i.i.d variables obeying the uniform distribution over $[\pi/3,2\pi/3]$. 
The channel complex gains follow $g_{1,k}\sim \mathcal{CN}(0,0.9)$ and $g_{l,k}\sim \mathcal{CN}(0,0.1)$ for $l > 1$. 
The CEEs follow i.i.d zero mean CSCG distribution with the same normalized mean square error (NMSE), i.e., $\zeta^2=\mathbb{E}[|\mathbf{H}_k-\hat{\mathbf{H}}_k|^2]/\mathbb{E}[|\mathbf{H}_k|^2],\forall k$, and then $\mathbf{A}_k=\mathbf{I}$, $\mathbf{B}_k=\zeta^2 \cdot \mathbf{I}$ \cite{CEEs}.
The learning rate and reward discount factor of the proposed MADDPG are 0.01 and 0.95, respectively. Unless specifically stated otherwise, our policies are characterized by a two-layer rectified linear unit (ReLU) multi-layer perception (MLP) with 64 units per layer \cite{MADDPG}. The signal-to-noise ratio (SNR) and the normalized region size $A/\lambda$ are set 30 dB and 3 unless otherwise stated.

We show the reward and sum-rate performance versus training time slots for the proposed MADDPG Algorithm for MA-enabled MIMO systems and the benchmark schemes in Fig.~\ref{fig3}. 
% To show clearly, we smoothed the reward and sum-rate.
The FPA system only needs to optimize the transmit beamforming vector. In comparison,  the proposed heterogeneous MADDPG and TR-MADDPG need to simultaneously optimize the transmit beamforming vector and DS antenna movement, which require more training time slots for convergence. Furthermore, it can also be observed that the proposed heterogeneous MADDPG algorithm can significantly improve the sum-rate by optimizing the DS antenna movement. However, due to the aforementioned limitations, the performance of the TR-MADDPG algorithm does not differ significantly from the FPA method.

Fig.~\ref{fig4} illustrates the sum-rate versus the normalized region size under perfect CSI and imperfect CSI. Several benchmarks are provided with the proposed \textbf{Algorithm 1}: 1) \textbf{FPA}.
% : The transmitter and $k$-th receiver are equipped with $N$ and $M_k$ fixed position antennas, spaced by $\lambda/2$ \cite{TWC-capacity}. 
2) \textbf{RMA}: The transmitter and the $k$-th receiver are equipped with $N$ FPAs and $M_k$ MAs \cite{TWC-capacity}. 3) \textbf{TR-MADDPG}. In both perfect and imperfect CSI scenarios, it can be observed that DS antenna movement optimization can achieve significant performance improvements compared to RMA and FPA due to its utilization of DS DoFs for attaining more favorable channels. Additionally, the sum-rate gain increases with the normalized region size and almost converges when the normalized region size is larger
than 3. Furthermore, the heterogeneous MADDPG algorithm achieves a higher sum-rate as compared to TR-MADDPG as the heterogeneous MADDPG framework separates the learning of the policies of beamforming and MA movement, thereby avoiding the outdated CSI. Moreover, it is evident that the presence of CEEs significantly affects system performance, with larger CEEs leading to more severe performance degradation, indicating the necessity of the propoer channel estimation method.

Fig.~\ref{fig5} shows the sum-rate performance of different schemes under perfect CSI and imperfect CSI versus SNR. It can be observed that with the same SNR, the MA-enabled system and the proposed algorithm can achieve a higher sum-rate as compared to benchmarks. For the case with SNR = 30 dB, the proposed algorithm has $22.18\%$, $47.97\%$ and $49.87\%$ sum-rate improvements over the RMA, FPA and TR-MADDPG schemes, respectively.

\section{Conclusion}
This paper investigated an MA-enabled multi-receiver communication system. To maximize the sum-rate of all receivers under imperfect CSI, we proposed a heterogeneous MADDPG algorithm which integrates two types of agents, namely the beamforming agent and the MA agent, to learn policies for beamforming and movement of MAs, respectively. Based on the offline learning under numerous imperfect CSI, the proposed MADDPG can output robust solutions for transmit beamforming and antenna movement in real time. Simulation results validate that the proposed algorithm for the MA-enabled system can achieve a significant sum-rate performance improvement compared with benchmarks and the sum-rate gain increases with the SNR and the normalized region size.

% \bibliographystyle{IEEEtran}
% \nocite{*}
% \bibColoredItems{blue}{chen-twc} 
% \bibColoredItems{blue}{guo-tcom} 
% \bibliography{WCL} 
\begin{comment}

\end{comment}

\bibliographystyle{IEEEtran}
\bibliography{WCL}

\begin{thebibliography}{10}
\providecommand{\url}[1]{#1}
\csname url@samestyle\endcsname
\providecommand{\newblock}{\relax}
\providecommand{\bibinfo}[2]{#2}
\providecommand{\BIBentrySTDinterwordspacing}{\spaceskip=0pt\relax}
\providecommand{\BIBentryALTinterwordstretchfactor}{4}
\providecommand{\BIBentryALTinterwordspacing}{\spaceskip=\fontdimen2\font plus
\BIBentryALTinterwordstretchfactor\fontdimen3\font minus
  \fontdimen4\font\relax}
\providecommand{\BIBforeignlanguage}[2]{{%
\expandafter\ifx\csname l@#1\endcsname\relax
\typeout{** WARNING: IEEEtran.bst: No hyphenation pattern has been}%
\typeout{** loaded for the language `#1'. Using the pattern for}%
\typeout{** the default language instead.}%
\else
\language=\csname l@#1\endcsname
\fi
#2}}
\providecommand{\BIBdecl}{\relax}
\BIBdecl

\bibitem{mimo}
M.~Z. Chowdhury, M.~Shahjalal, S.~Ahmed, and Y.~M. Jang, ``{6G} wireless
  communication systems: Applications, requirements, technologies, challenges,
  and research directions,'' \emph{IEEE Open J. Commun. Soc.}, vol.~1, pp.
  957--975, Jul. 2020.

\bibitem{AS}
M.~Gharavi-Alkhansari and A.~B. Gershman, ``Fast antenna subset selection in
  {MIMO} systems,'' \emph{IEEE Trans. Signal Process.}, vol.~52, no.~2, pp.
  339--347, Feb. 2004.

\bibitem{FAS}
K.-K. Wong, A.~Shojaeifard, K.-F. Tong, and Y.~Zhang, ``Fluid antenna
  systems,'' \emph{IEEE Trans. Wireless Commun.}, vol.~20, no.~3, pp.
  1950--1962, Mar. 2021.

\bibitem{MA-Mag}
L.~Zhu, W.~Ma, and R.~Zhang, ``Movable antennas for wireless communication:
  Opportunities and challenges,'' \emph{IEEE Commun. Mag.}, to appear, 2023.

\bibitem{TWC-2}
------, ``Modeling and performance analysis for movable antenna enabled
  wireless communications,'' \emph{IEEE Trans. Wireless Commun.}, to appear,
  2023.

\bibitem{TWC-capacity}
W.~Ma, L.~Zhu, and R.~Zhang, ``Mimo capacity characterization for movable
  antenna systems,'' \emph{IEEE Trans. Wireless Commun.}, 2023, to appear.

\bibitem{multiuser-1}
L.~Zhu, W.~Ma, B.~Ning, and R.~Zhang, ``Movable-antenna enhanced multiuser
  communication via antenna position optimization,'' \emph{IEEE Trans. Wireless
  Commun.}, 2023, to appear.

\bibitem{multiuser-2}
Y.~Wu, D.~Xu, D.~W.~K. Ng, W.~Gerstacker, and R.~Schober, ``Movable
  antenna-enhanced multiuser communication: Jointly optimal discrete antenna
  positioning and beamforming,'' in \emph{Proc. IEEE Global Common. Conf.},
  Kuala Lumpur, Malaysia, Dec. 2023, pp. 7508--7513.

\bibitem{multiuser-3}
S.~Yang, W.~Lyu, B.~Ning, Z.~Zhang, and C.~Yuen, ``Flexible precoding for
  multi-user movable antenna communications,'' \emph{IEEE Wireless Commun.
  Lett.}, to appear, 2024.

\bibitem{multiuser-4}
G.~Hu, Q.~Wu, K.~Xu, J.~Ouyang, J.~Si, Y.~Cai, and N.~Al-Dhahir,
  ``Movable-antenna array enabled multiuser uplink: A low-complexity gradient
  descent for total transmit power minimization,'' arXiv preprint: 2312.05763,
  2023.

\bibitem{single-stream}
C.~Zhong, T.~Ratnarajah, S.~Jin, and K.-K. Wong, ``Performance analysis of
  optimal single stream beamforming in mimo dual-hop af systems,'' \emph{IEEE
  J. Sel. Areas Commun.}, vol.~30, no.~8, pp. 1415--1427, Sep. 2012.

\bibitem{CE}
W.~Ma, L.~Zhu, and R.~Zhang, ``Compressed sensing based channel estimation for
  movable antenna communications,'' \emph{IEEE Commun. Lett.}, vol.~27, no.~10,
  pp. 2747--2751, Oct. 2023.

\bibitem{CEEs-2}
P.~Zeng, D.~Qiao, H.~Qian, and Q.~Wu, ``Joint beamforming design for {IRS}
  aided multiuser {MIMO} with imperfect {CSI},'' \emph{IEEE Trans. Veh.
  Technol.}, vol.~71, no.~10, pp. 10\,729--10\,743, Oct. 2022.

\bibitem{CEEs}
J.~Yaswanth, M.~Katwe, K.~Singh, S.~Prakriya, and C.~Pan, ``Robust beamforming
  design for active-{RIS aided MIMO SWIPT} communication system: A power
  minimization approach,'' \emph{IEEE Trans. Wireless Commun.}, 2023, to
  appear.

\bibitem{math}
E.~Jorswieck and H.~Boche, ``Majorization and matrix-monotone functions in
  wireless communications,'' \emph{Found. Trends Commun. Inf. Theory}, vol.~3,
  no.~6, pp. 553--701, 2006.

\bibitem{MADDPG}
R.~Lowe, Y.~Wu, A.~Tamar, J.~Harb, P.~Abbeel, and I.~Mordatch, ``Multi-agent
  actor-critic for mixed cooperative-competitive environments,'' in \emph{Proc.
  Adv. Neural Inf. Process. Syst. (NeurIPS)}, 2017, pp. 6382--6393.

\end{thebibliography}

\end{document}